\documentstyle[epsfig,longtable]{aipproc}

\begin{document}

\title{KiloHertz Quasi-Periodic Oscillations in the Z sources GX
340$+$0, Cygnus X-2, GX 17$+$2, GX 5$-$1, and Scorpius X-1}
 
\author{Rudy Wijnands \& Michiel van der Klis}
\address{Astronomical Institute ``Anton Pannekoek''\\University of
Amsterdam\\Kruislaan 403, NL-1098 SJ, Amsterdam\\The Netherlands}

\maketitle

\begin{abstract}We have discovered kiloHertz quasi-periodic oscillations 
(kHz QPOs) in five Z sources: GX 340$+$0, Cygnus X-2, GX 17$+$2, GX
5$-$1, and Scorpius X-1. In all sources the properties of these kHz
QPOs are very similar and closely related to the position of the
sources on the Z track traced out in the X-ray color-color diagram and
the hardness-intensity diagram. The frequencies of the kHz QPOs
increase when the sources move from the left end of the horizontal
branch to horizontal/normal branch vertex, thus with inferred mass
accretion rate. Only for Scorpius X-1 the kHz QPOs have been observed
down the normal branch unto the flaring branch.  The strength and the
FWHM of the higher-frequency kHz QPOs decrease with mass accretion
rate, but when the lower-frequency kHz QPOs are detected the strength
and the FWHM of this QPO stay approximately constant with mass
accretion rate.  In {Scorpius\,X-1} the frequency separation between
the kHz QPOs decreases with mass accretion rate, but in the other Z
sources the separation remains approximately constant, although a
similar decrease in peak separation as found in Scorpius X-1 can not
be excluded.

\end{abstract}

\section*{Introduction}

KiloHertz quasi-periodic oscillations (kHz QPOs) have been observed
for more then a dozen low-mass X-ray binaries (LMXBs) with the Rossi
X-ray Timing Explorer (RXTE). Five of those LMXBs are part of the
subclass called Z sources \cite{hk89}. Z sources trace out a Z like
shape in the X-ray color-color diagram (CD) and hardness-intensity
diagram (HID). The branches are called horizontal branch (HB), normal
branch (NB), and flaring branch (FB). The mass accretion rate is
thought to increase from the HB, via the NB, to the FB \cite{hk89}.
The timing behavior below $\sim$100 Hz is closely related to the
position of the source on this Z track (see van der Klis 1995
\cite{klis95} for a review).  In this paper we summarize the results
of the properties of the kHz QPOs in Z sources and we will show that
the properties of these kHz QPOs are also related to the position of
the sources on the Z track.

\section*{Observations and analysis}

We observed all Z sources on several occasions with the proportional
counter array (PCA) onboard RXTE. For the details of the observations
we refer to van der Klis et al. (1996 \cite{klis96}; 1997
\cite{klis97}; Scorpius X-1), Wijnands et al. (1997 \cite{wijnands97};
1998a \cite{wijnands98a}; 1998b \cite{wijnands98b}; GX 17$+$2, Cygnus
X-2, and GX 5$-$1), and Jonker et al. (1998 \cite{jonker}; GX
340$+$0). On several occasions only 3 or 4 out of the five PCA
detectors were on. For constructing the CDs and HIDs we used only the
data of the three detectors which were always on. In our power
spectrum analysis we used all data available. Power spectra were
selected by the position of the sources on the their Z tracks.

\section*{Results and discussion}

\begin{figure}[t]
\centerline{\epsfig{file=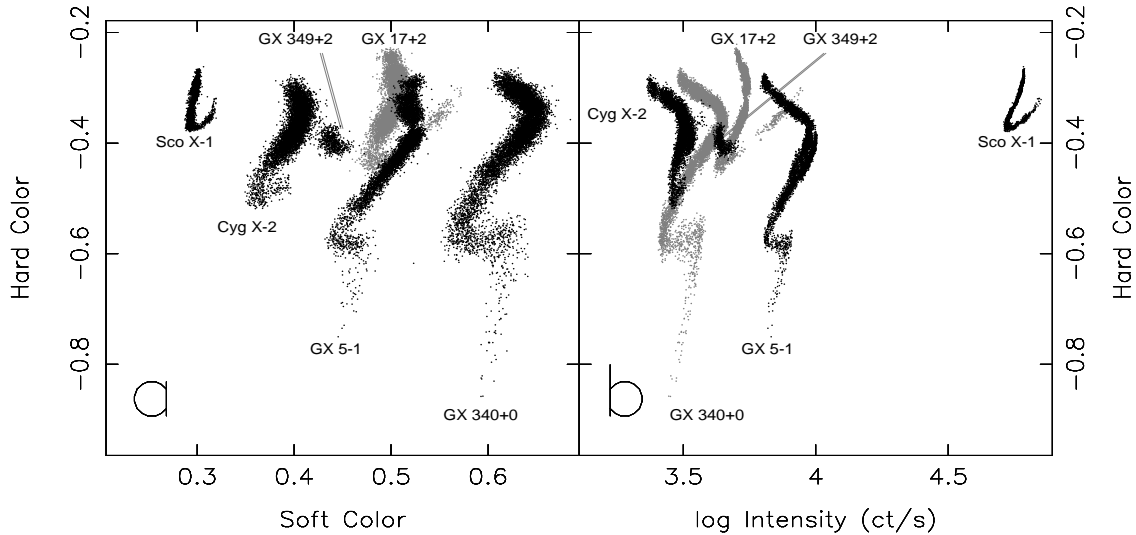,width=15cm,height=7.5cm}}
\vspace{5pt}
\caption{Color-color diagrams ({\it a}) and hardness-intensity
diagrams ({\it b}) of the Z sources GX\,349$+$2, GX\,340$+$0,
Cygnus\,X-2, GX\,5$-$1, GX\,17$+$2, and Scorpius X-1. The soft color is
the logarithm of the count rate ratio between 3.5--6.4 keV and
2.0--3.5 keV; the hard color is the logarithm of the count rate ratio
between 9.7--16.0 keV and 6.4--9.7 keV; the intensity is the logarithm
of the 3-detector count rate in the photon energy range 2.0--16.0
keV. Both diagrams are corrected for background ($\sim$50 ct/s in the
energy range 2.0--16.0 keV); the count rate is not dead-time
corrected. All points are 16 s averages. Some Z tracks are plotted in
gray for clarity.}\label{CD}
\end{figure}

The X-ray color-color diagrams and hardness-intensity diagrams of the
Z sources are shown in Figures \ref{CD}a and \ref{CD}b,
respectively. Also the data of the Z source GX\,349$+$2, which has
only been observed for a short period of time (see Kuulkers \& van der
Klis 1997 \cite{kuulkers}), has been included.  For Scorpius X-1 we
have only plotted a very small part of the total available data. The
spectral data of Scorpius X-1 are affected by instrumental effects
mainly due to different offset angles used at the time of the
observations. We only plotted data with similar offset angles.

We observe a clear Z track for all the Z sources, except for
GX\,349$+$2, showing all three branches.  The Z tracks of the two Z
sources GX\,5$-$1 and GX\,340$+$0 are very similar in shape. Both
sources trace out a clear Z track and an extra branch is available at
the end of the FB. Such behavior has already been reported before in
GX 5$-$1 \cite{kuulkers94} and GX 340$+$0 \cite{penninx91}.

\begin{figure}[t]
\centerline{\epsfig{file=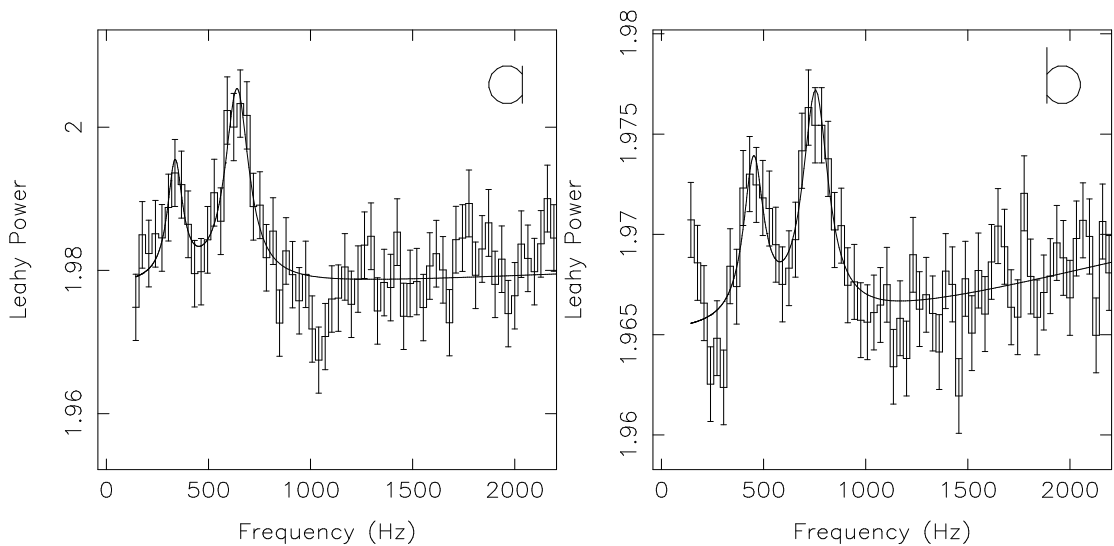,width=15cm,height=7.5cm}}
\vspace{5pt}
\caption{Typical Leahy normalized power density spectra of GX 5$-$1
({\it a}) and GX 340$-$0 ({\it b}) showing the kHz QPOs. The upwards
slope at kHz frequencies, especially in {\it b}, is due to
instrumental dead time. }\label{QPO}
\end{figure}

In all the Z sources, except GX 349+2 \cite{kuulkers}, we discovered
kHz QPOs with frequencies between 300 and 1100 Hz on the horizontal
branch
\cite{klis96,klis97,wijnands97,wijnands98a,wijnands98b,jonker}. Typical
power density spectra of GX 5$-$1 and GX 340$+$0 showing the kHz QPOs
in these sources are shown in Figure \ref{QPO}. Figures of the kHz
QPOs in the other sources can be found elsewhere (Scorpius X-1:
\cite{klis96,klis97}; GX 17$+$2: \cite{wijnands97}; Cygnus X-2:
\cite{wijnands98a}). The only source for which the kHz QPOs are also
observed on the (lower) normal branch and the flaring branch is
Scorpius X-1 \cite{klis96,klis97}.

In all the Z sources for which kHz QPOs have been discovered, the
properties of the kHz QPOs as a function of the position of the
sources on the Z track are very similar and are strongly related to
the position of the source on the Z track. Both kHz QPOs increase in
frequency when the inferred mass accretion rate increases. The rms
amplitude and the FWHM of the higher-frequency kHz QPO decreases with
mass accretion rate, but the rms amplitude and FWHM of the
lower-frequency kHz QPO stays approximately constant when this QPO is
detected.  Typically the rms amplitude of the higher-frequency kHz QPO
decreases from 5--6\% (in the energy range 5.0--60 keV) at the left end
of the HB to $\sim$2\% at the HB/NB vertex, whereas the
lower-frequency kHz QPOs remains approximately constant at 2--4\%
(depending on the source).  In {Scorpius\,X-1} the frequency
separation between the kHz QPOs decreases with mass accretion rate
\cite{klis97}, but in the other Z sources the separation remains
approximately constant (although a similar decrease in peak separation
as found in Scorpius X-1 can not be excluded).

Simultaneous with the kHz QPOs in all the Z sources displaying kHz QPO
behavior, the horizontal branch QPO and its second harmonic are
observed. This rules out the possibility that both types of QPOs are
caused by the same physical mechanism. Recently, Stella \& Vietri
(1998 \cite{stella}) proposed that the low frequency QPOs observed in
several less luminous LMXBs are due to a precession of the innermost
disk region, dominated by the Lense-Thirring effect. Following their
reasoning and assumptions, and assuming that the neutron star spin
frequency is the kHz QPO separation (290--360 Hz, depending on the
source), we derive maximum precession frequency significantly (about a
factor 2) below the observed frequencies of the HBOs in the Z
sources. For Lense-Thirring precession to explain the HBO one would
have to postulate that the observed HBO frequencies are those of the
second harmonic of the precession frequency, or that the neutron star
spin frequency is {\it twice} the peak separation.

\end{document}